\documentclass[prd,preprint,nofootinbib]{revtex4}

\usepackage{graphicx, cancel}

\begin{document}

\newcommand{\gsim}{ \mathop{}_{\textstyle \sim}^{\textstyle >} }
\newcommand{\lsim}{ \mathop{}_{\textstyle \sim}^{\textstyle <} }


\renewcommand{\thefootnote}{\fnsymbol{footnote}}
\preprint{DESY 08-051}
\title{Lepton-flavour violation \\
in the light of leptogenesis and muon $g-2$}
\renewcommand{\thefootnote}{\alph{footnote}}

\author{Motoi Endo and Tetsuo Shindou}

\affiliation{
Deutsches Elektronen Synchrotron DESY, Notkestrasse 85,
22607 Hamburg, Germany\\
}

\begin{abstract}
\noindent
A discrepancy between the experimental and theoretical results was 
reported for the anomalous magnetic moment of the muon. 
In leptogenesis scenarios which are due to a decay of right-handed 
neutrinos, this anomaly leads to large violations of lepton flavour 
within the framework of supersymmetric see-saw models. It is shown 
that for a hierarchical right-handed neutrino mass spectrum, 
we generically expect to observe $\mu \to e \gamma$ in the near 
future experiment. 
\end{abstract}

\maketitle

\section{Introduction}

The experiments have measured the neutrino oscillations and 
established a mass of neutrinos. This cannot be explained in 
the Standard Model (SM) framework and thus is an evidence 
of physics beyond SM. From the cosmological side, one of the 
most challenging mysteries of the universe is the source of the 
asymmetry of the baryon number density. Since it is quite 
difficult to explain the asymmetry in SM, we are required to 
introduce new mechanism to generate it. 

Recently, the anomalous magnetic moment $(g-2)$ of the muon 
suggested a contribution of the new physics. The E821 experiment 
at Brookhaven measured it at the extremely precise level\cite{Bennett:2006fi}, 
and the results have been compared with the SM prediction. 
According to the latest update of the hadronic 
contribution to the SM value\cite{Hagiwara:2006jt}, which is based 
on the $e^+e^-$ collision data, a difference was reported 
between these two results as (see \cite{Passera:2008jk} and references 
therein for details)
\begin{eqnarray}
  a_\mu({\rm exp.}) - a_\mu({\rm SM}) \;=\; 302(88) \times 10^{-11},
  \label{eq:muong-2}
\end{eqnarray}
which means 3.4$\sigma$ deviation. One may expect that this 
discrepancy is explained by hadronic uncertainties in the 
SM prediction. However, too large corrections are then required 
to fix (\ref{eq:muong-2}) in the hadron sector\cite{Passera:2008jk}. 
Instead of pursuing this idea, we consider new contributions from 
physics beyond SM. 

These features are easily achieved by the supersymmetric (SUSY) 
see-saw scenarios. In the SUSY models, all the SM particles are 
accompanied by their superpartners, and these new particles affect 
low-scale phenomena through radiative corrections. In fact, the 
discrepancy of the muon $g-2$ can be saturated due to a $\tan\beta$ 
enhancement\cite{SUSYg-2}. On the other hand, the neutrino oscillations 
are realized by the Type I see-saw mechanism, introducing the 
right-handed sector\cite{SEESAW}. Furthermore, against the mystery 
of the baryon asymmetry, the see-saw mechanism provides an elegant 
solution. By producing the right-handed neutrinos at an early stage of 
the universe, the lepton asymmetry would be generated at the decay 
of those neutrinos, and then leads to the baryon asymmetry of Universe 
(BAU) via the sphaleron effect. 

As a prediction of the SUSY see-saw scenarios, the lepton flavours 
are violated at the weak scale. These violations are tightly correlated 
with the leptogenesis and muon $g-2$. The measurements of the 
neutrino oscillations indicate flavour-violations in the neutrino sector. 
Such violations are naturally transmitted into the charged-lepton 
sector via the renormalization group evolutions. At the weak scale, 
the lepton-flavour violating processes are induced by superparticles. 
Those superparticles contribute as well to the prediction of the muon 
$g-2$. Thus, they cannot be arbitrary heavy in order to explain the 
muon $g-2$ anomaly (\ref{eq:muong-2}). On the other hand, the 
right-handed neutrino mass scale is required to be large for the 
leptogenesis to work successfully. Such a large scale enhances the 
neutrino Yukawa coupling including the flavour-violating effects. 
Consequently, in the leptogenesis scenarios, it is considered that 
the lepton-flavour violating processes tend to be sizable in the light 
of the anomaly of the muon $g-2$. In this letter, we will show that 
for a hierarchical right-handed neutrino mass spectrum, $\mu \to 
e \gamma$ is generically expected to be observed in the near 
future experiment.
\section{Correlations Among Observables}
\label{sec:Sec2}

We first briefly review the muon $g-2$. The SUSY contributions 
consist of the two diagrams; those mediated by the charginos and 
neutralinos. Considering the coupling strength, the former usually 
dominates the SUSY contributions. The result is approximately 
obtained as~\cite{SUSYg-2}
\begin{eqnarray}
  \delta a_\mu \;\simeq\; \frac{5\alpha_2}{48\pi}
  \frac{m_{\mu}^2}{m_{\text{SUSY}}^2}\mathrm{sign}(M_2\mu_H)
  \tan\beta\;,
\label{eq:g-2MI}
\end{eqnarray}
where $M_2$ is a wino mass, $\mu_H$ is a Higgsino mass and
$\tan\beta=\langle H_u\rangle/\langle H_d\rangle$ is a ratio of two 
vacuum expectation values of Higgses. In this expression, we have 
set all soft masses to be equal for simplicity. In the letter, we choose 
a sign of $(M_2 \mu_H)$ to be positive to explain (\ref{eq:muong-2}), 
which is irrelevant for the following two observables. 

The second observable is ${\rm Br}(\mu \to e \gamma)$. 
The SM contribution is known to be highly suppressed, and the 
SUSY effects determines the branching ratio. The SUSY diagrams 
are the same as those of the muon $g-2$ except for the flavour 
dependence. Let us consider that the soft breaking slepton masses 
are flavour universal at the cutoff scale, $\mu_X$, which is assumed 
to be larger than the right-handed neutrino mass scale, $\mu_R$. 
Even with this flavour-universal condition, in SUSY see-saw models, 
neutrino Yukawa couplings induce the left-handed slepton mixing 
through the renormalization running from $\mu_X$ down to $\mu_R$. 
Thus we expect significant SUSY contributions to ${\rm Br}(\mu\to 
e\gamma)$. With the universal slepton mass boundary condition, 
the branching ratio is represented as\cite{SUSYmeg}
\begin{eqnarray}
  {\rm Br}(\mu \to e \gamma) \;\sim\; 
\frac{\alpha^3}{G_F^2}
\frac{\left|(Y_N^{\dagger}Y_N)_{12}\ln\left(\mu_R/\mu_X\right)\right|^2}
{m_{\rm SUSY}^4}\tan^2\beta\;,
\label{eq:BrmegMI}
\end{eqnarray}
up to a numerical factor, where $Y_N$ is a neutrino Yukawa coupling 
matrix in the basis of diagonal charge lepton Yukawa couplings and 
right-handed neutrino mass matrix. Here, all the soft SUSY breaking 
parameters including those during the renormalization running were 
simply set to be equal. 
Comparing with Eq.~(\ref{eq:g-2MI}), a naive relation between 
$\delta a_{\mu}$ and ${\rm Br}(\mu\to e\gamma)$ is found as 
${\rm Br}(\mu\to e\gamma)\propto (\delta a_{\mu})^2$ for fixed 
$Y_N$\cite{HisanoTobe}. 

It should be mentioned that we showed Eqs.~(\ref{eq:g-2MI}) and 
(\ref{eq:BrmegMI}) to clarify the dependence on the model parameters, 
while for the following numerical analysis, we do not rely on these 
simplifications. Namely, we will evaluate $\delta a_{\mu}$ and 
${\rm Br}(\mu\to e\gamma)$ completely up to the one-loop level 
after solving a set of renormalization group equations and diagonalizing 
the mass matrix in the next section.

The production of the right-handed neutrino leads to the generation 
of the lepton asymmetry. In leptogenesis scenarios such as the thermal 
and non-thermal ones, the lepton asymmetry is proportional to 
CP asymmetry in a decay of a lightest right-handed neutrino, $N_1$, 
into lepton doublet, $L$, and Higgs, $H_u$. The CP asymmetry parameter 
$\epsilon_1$ is  (see \cite{Buchmuller:2004nz} for a review)
\begin{eqnarray}
\epsilon_1\;&\equiv&\;
\frac{\Gamma(N_1\to L+H_u)-\Gamma(N_1\to L^c+H_u^c)}
{\Gamma(N_1\to L+H_u)+\Gamma(N_1\to L^c+H_u^c)}\nonumber\\
&=&\frac{1}{8\pi(Y_NY_N^{\dagger})_{11}}
\sum_{i\neq 1}\mathrm{Im}\left[
(Y_NY_N^{\dagger})_{i1}^2f(M_i^2/M_1^2)
\right]\;,
\label{eq:formula-epsCP}
\end{eqnarray}
where $M_i$'s are masses of the right-handed neutrinos and 
a loop function $f(x)$ is given as
\begin{equation}
f(x) = \sqrt{x}\ln\left(1+\frac{1}{x}\right)-\frac{2\sqrt{x}}{x-1}\;.
\end{equation}
Here and in the following analysis, we ignore flavour-dependent effects 
on the leptogenesis\cite{FlavourLeptogenesis} because they are generically 
negligible. The above expression is valid for a hierarchical right-handed neutrino 
mass spectrum, {\it i.e.}, $|M_{2,3}-M_1|\gg \Gamma_{2,3}+\Gamma_1$, 
where $\Gamma_i$ is the decay width of the $i$th right-handed neutrino 
and estimated at the tree level as
\begin{equation}
\Gamma_i = \frac{(Y_NY_N^{\dagger})_{ii}}{8\pi}M_i\;.
\end{equation}
When $M_2$ becomes very close to $M_1$, the resonant effects 
contribute to the lepton asymmetry. This topic will be discussed 
in the final section. In the following analysis, we will focus on 
the hierarchical (off-resonant) case, $M_1\ll M_2 \lsim M_3$. 

The neutrino Yukawa coupling matrix and the right-handed neutrino 
mass matrix are connected to the light neutrino mass matrix, $m_{\nu}$ 
through the see-saw relation,
\begin{equation}
(Y_N)_{ki}\frac{1}{M_k}(Y_N)_{kj}\langle H_u\rangle^2=(m_{\nu})_{ij}
=U^*_{ik}m_k U^*_{jk}\;,
\label{eq:seesaw}
\end{equation}
where $m_i$ is mass eigenvalues of the light neutrinos and 
$U$ is Pontecorvo--Maki--Nakagawa--Sakata (PMNS) matrix\cite{PMNS}.
From data of neutrino oscillation experiments, 
parameters in $U$ and the squared 
mass differences are determined as\cite{NuEXP}
\begin{eqnarray}
|\Delta m_{\rm atm}^2|&\equiv& |m_3^2-m_1^2|\sim 2.5\times 10^{-3}\text{eV}^2\;,
\quad
\sin^22\theta_{\rm atm}= 4|U_{23}^2|(1-|U_{23}^2|)\sim 1.0\;,\nonumber\\
\Delta m_{\odot}^2&\equiv& m_2^2-m_1^2\sim 8.0\times 10^{-5}\text{eV}^2\;,
\quad
\tan^2\theta_{\odot}\simeq \frac{|U_{12}^2|}{|U_{11}^2|}\sim 0.4\;,\\
|U_{13}|&\lesssim& 0.2\;.\nonumber
\label{eq:neutrino-data}
\end{eqnarray}
In order to incorporate the relation Eq.~(\ref{eq:seesaw}) in the analysis, 
it is useful to use the following parametrization\cite{CasasIbarra},
\begin{equation}
(Y_N)_{ij} = \frac{1}{\langle H_u\rangle}\sqrt{M_i}R_{ik}\sqrt{m_k}U_{jk}^*\;,
\label{eq:R-param}
\end{equation}
with $R_{ij}$ satisfying $\sum_kR_{ik}R_{jk}=\delta_{ij}$. $R_{ij}$ has 
six real parameters.

In a lot of leptogenesis scenarios, the lepton asymmetry is favored to be 
as large as possible. With the above parametrization, it is easily shown 
that a size of the CP asymmetry parameter $\epsilon_1$ has an upper 
bound which is much less than one. With $M_1/M_{2,3}\ll 1$ and 
${\rm max}(|R_{ij}|) < \mathcal{O}(M_{2,3}/M_1)$ (see \cite{Raidal:2004vt}), 
one gets\cite{DavidsonIbarra}
\begin{eqnarray}
\left|\epsilon_1\right|&\simeq&
\frac{1}{8\pi(Y_NY_N^{\dagger})_{11}}
\left|
\sum_{i\neq 1}\mathrm{Im}\left(
(Y_NY_N^{\dagger})_{i1}^2\frac{M_1}{M_i}
\right)
\right|
=
\frac{M_1}{8\pi\langle H_u\rangle^2}
\left|
\frac{\sum_i\mathrm{Im}(m_i^2R_{1i}^2)}
{\sum_im_i|R_{1i}^2|}\right|\nonumber\\
&\leq&
\frac{M_1}{8\pi\langle H_u\rangle^2}\frac{\Delta m_{\rm atm}^2}{m_1+m_3}\;.
\label{eq:maxeps1}
\end{eqnarray}
One can find that $|\epsilon_1|$ is maximized when $R_{12}=0$ 
and $|\mathrm{Re}(R_{13})|=|\mathrm{Im}(R_{13})|$ are satisfied.
For a given maximal $\epsilon_1\equiv\epsilon_1^{\text{max}}$, the 
right-handed neutrino mass is estimated as
\begin{equation}
M_1
\simeq
1.5\times 10^{10}\text{GeV}
\left(\frac{|\epsilon_1^{\text{max}}|}{10^{-6}}\right)
\left(\frac{\langle H_u\rangle}{174\text{GeV}}\right)^2
\left(\frac{\Delta m_{\text{atm}}^2}{2.5\times 10^{-3}\text{eV}^2}\right)^{-1}
\left(\frac{m_1+m_3}{0.05\text{eV}}\right)\;.
\label{eq:M1forpes1max}
\end{equation}

We expect that a large asymmetry of the lepton number density 
leads to a sizable branching ratio of $\mu \to e \gamma$\cite{LFV-Lepto,LFV-Lepto2}. 
From Eq.~(\ref{eq:maxeps1}), $|\epsilon_1|$ is proportional to $M_1$, 
and thus the CP asymmetry, i.e, the lepton asymmetry, is 
enhanced for heavier $M_1$. Then, by satisfying the relation 
$M_1 \ll M_2 \lsim M_3$, heavier $M_1$ leads to larger elements 
of $Y_N$. As was shown in Eq.~(\ref{eq:BrmegMI}), large $Y_N$ 
enhances $\text{Br}(\mu\to e\gamma)$. Thus, for leptogenesis 
scenarios to successfully work, ${\rm Br}(\mu \to e\gamma)$ tends 
to be large. 

One may consider that the SUSY contributions to the $\mu \to 
e \gamma$ amplitude are suppressed by heavy superparticles. 
However, the anomaly of the muon $g-2$, Eq.~(\ref{eq:muong-2}), 
prohibits the particles to be decoupled as long as it is explained 
by the SUSY contributions. Especially since the branching ratio 
of $\mu \to e \gamma$, Eq.~(\ref{eq:BrmegMI}), is tightly correlated 
with the SUSY contribution to the muon $g-2$, Eq.~(\ref{eq:g-2MI}), 
it is very hard to suppress $\text{Br}(\mu\to e\gamma)$ with 
keeping both $\delta a_{\mu}$ and $\epsilon_1$ large.

We should mention specific cases in which $\text{Br}(\mu\to e
\gamma)$ is suppressed. First of all, the flavour structure of 
the Yukawa coupling potentially causes a cancellation in the 
decay amplitude. Actually, the amplitude is proportional to 
$(Y_N^{\dagger}Y_N)_{12} = (Y_N)_{11}^*(Y_N)_{12} + 
(Y_N)_{21}^*(Y_N)_{22} + (Y_N)_{31}^*(Y_N)_{32}$. Thus, 
an accidental cancellation may happen among these complex 
numbers with $\delta a_{\mu}$ and $\epsilon_1$ fixed. 
We might also obtain cancellations between the chargino and 
neutralino contributions, or by taking into account the initial 
flavour-changing components of the soft SUSY breaking 
parameters. These cases will be commented in the final section. 
In any case, the cancellations are considered to be accidental 
in general and thus regarded as a fine-tuning. 
As the second case, let us consider texture structures of the 
right-handed Yukawa coupling. Assigning texture zeros properly 
for the Yukawa coupling, $\text{Br}(\mu\to e\gamma)$ can be 
suppressed\cite{LFV-Lepto2}. Then, instead of $\mu \to e\gamma$, 
it is likely to observe other lepton flavour violating processes, 
e.g., $\tau \to \mu(e) \gamma$. Note that, since the light-neutrino 
mass spectrum is predicted to be specific in this framework, it is 
expected to identify such a case in future. In the following study, 
we will discuss a lower bound of $\text{Br}(\mu\to e\gamma)$ in 
generic conditions. Namely, we will assume no fine-tunings in the 
neutrino sector, and not include the specific texture setup.

\section{Analysis}

\begin{figure}
\begin{tabular}{cc}
\includegraphics[scale=1.0]{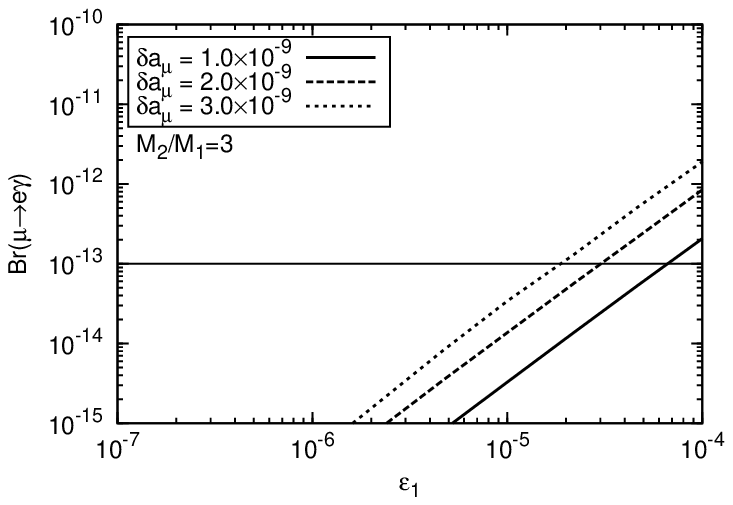}&
\includegraphics[scale=1.0]{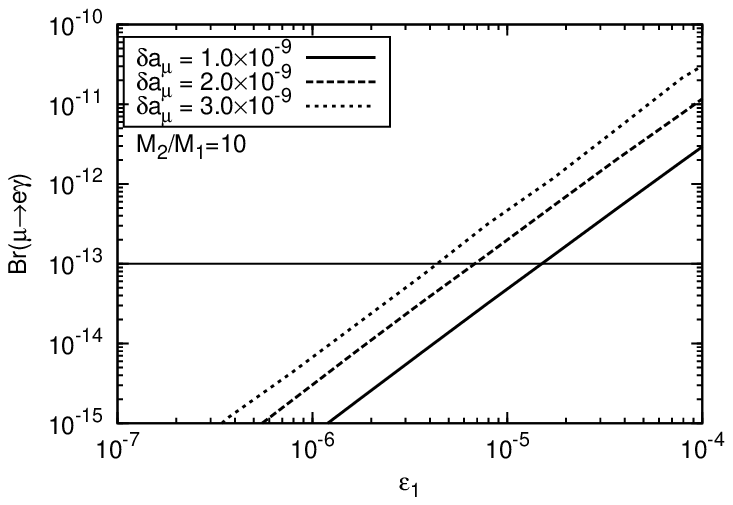}\\
(a)&(b)\\
\includegraphics[scale=1.0]{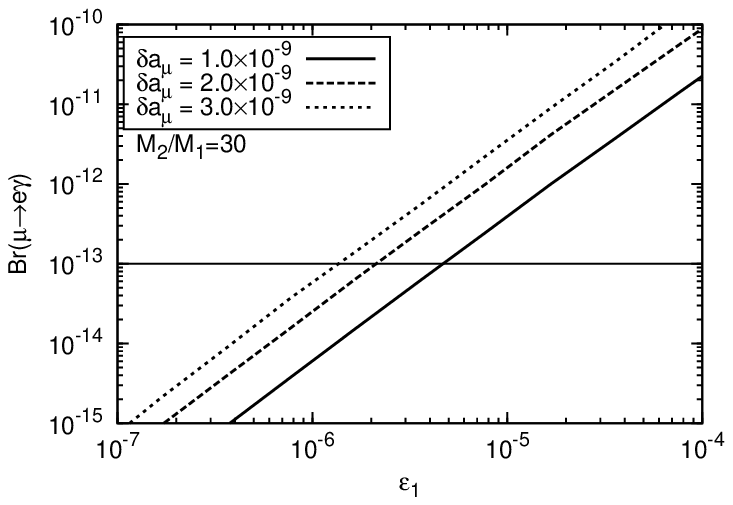}&
\includegraphics[scale=1.0]{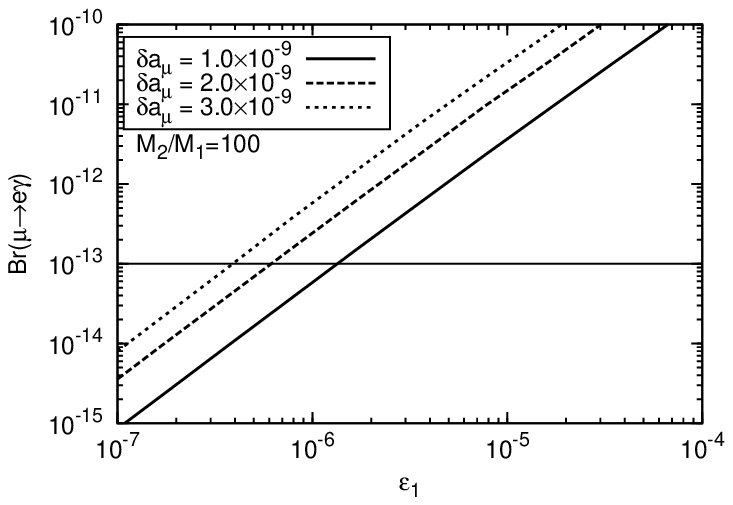}\\
(c)&(d)\\
\end{tabular}
\caption{Generic lower bound on the prediction of 
$\text{Br}(\mu\to e\gamma)$ for fixed $\delta a_{\mu}$ 
and $\epsilon_1$. 
Here, $\mu_X$ is set to be the GUT scale. 
In the analysis, we assumed no fine-tunings and did not 
take specific texture structures so that we avoid accidental 
suppression of $\text{Br}(\mu\to e\gamma)$. 
We also neglect the resonant effects for the leptogenesis, 
which become significant when $M_2$ is very close to $M_1$. 
In the graphs, the ratio $M_2/M_1$ is taken as 
(a) $M_2/M_1=3$, (b) $= 10$, (c) $= 30$, and $= 100$ 
with $M_1$ varied. The horizontal line at $\text{Br}(\mu\to 
e\gamma) = 10^{-13}$ represents the sensitivity 
of the near future experiment. 
}
\label{fig:epsCP-Brmeg}
\end{figure}

Lower bounds of the predicted $\text{Br}(\mu\to e \gamma)$ are 
displayed in Fig.~\ref{fig:epsCP-Brmeg} as a function of $\epsilon_1$ 
for fixed $\delta a_{\mu}$. Here, we scanned the SUSY parameters~\footnote{
  In the analysis, we set the soft mass of the selectron to be degenerate 
  with that of the smuon. When the selecton is much heavier than the 
  smuon, $\text{Br}(\mu\to e \gamma)$ can be suppressed for fixed 
  $\delta a_\mu$.
}
and the neutrino Yukawa couplings satisfying the experimental data 
(\ref{eq:neutrino-data}). 
In the analysis, we set $\mu_X$ to be the grand-unification (GUT) 
scale and imposed the following assumptions; neglecting resonant 
effects of the leptogenesis and generic structure for the lepton 
sector. In particular, as for the latter assumption, we assumed no 
fine-tunings in the right-handed neutrino Yukawa coupling and 
sought the parameter points which minimize 
$|(Y_N)_{11}^*(Y_N)_{12}|+|(Y_N)_{21}^*(Y_N)_{22}| + 
|(Y_N)_{31}^*(Y_N)_{32}|$ for fixed $\epsilon_1$. This means that 
$|(Y_N^{\dagger}Y_N)_{12}|$ is minimized without an accidental 
cancellation among three different terms in the summation on the lines. 

It can be said that the lower bounds in Fig.~\ref{fig:epsCP-Brmeg} 
are conservative under the above assumptions. We can check that 
on the lines, $m_1 = 0$ and $U_{13}=0$ are satisfied, and 
$|\epsilon_1|$ is maximized by satisfying $R_{12}=0$ and 
$\arg(R_{13})=\pi/4$. When we increase the lightest neutrino mass, 
according to Eq.~(\ref{eq:maxeps1}) the maximal value of $\epsilon_1$ 
is suppressed by $\Delta m_{\text{atm}}^2/(m_1 + m_3)$. Namely, with 
$\epsilon_1$ fixed, $M_1$ increases for larger $m_1$, and thus, the 
lower bound of $\text{Br}(\mu\to e\gamma)$ goes up. In addition, 
$\text{Br}(\mu\to e\gamma)$ is minimized when both $m_1=0$ 
and $U_{13}=0$ are satisfied. This is because $(Y_N)_{11}$ becomes 
naturally small. From (\ref{eq:R-param}), if either $m_1$ or $U_{13}$ 
is finite, the branching ratio of $\mu\to e\gamma$ receives an additional 
contribution from $(Y_N)_{11}^*(Y_N)_{12}$. Also, the CP violation 
phases in the PMNS matrix are taken to be zero, otherwise $\text{Br}
(\mu \to e\gamma)$ becomes larger. Although we assumed that the slepton 
mass matrix is universal at the cutoff scale, introducing the off-diagonal 
components is just additive to the branching ratio. In conclusion, 
the lower bounds in Fig.~\ref{fig:epsCP-Brmeg} are conservative. 

It is stressed that the lower bound of $\text{Br}(\mu\to e\gamma)$
depends on $M_2/M_1$ but is independent of $M_3$. 
It is because on the lines of the lower bound, $R_{ij}$ still has 
enough degrees of freedom to give the same minimal value of 
$\text{Br}(\mu\to e\gamma)$ for different values of $M_3/M_2$. 
In other words, the minimum of $\text{Br}(\mu\to e\gamma)$ has 
a flat direction in the parameter space. It is obtained that as long as 
$M_2/M_1$ is larger than 10, the lower bound is propotional to 
$M_2/M_1$.While for smaller $M_2/M_1$, the lower bound of 
$\text{Br}(\mu\to e\gamma)$ is not a linear function of $M_2/M_1$, 
because contributions from the next leading order in loop function 
$f(M_i^2/M_1^2)$ which is a term of order $M_1^3/M_2^3$ are 
non-negligeble. On the other hand, the lower bound is almost  
independent of $\tan\beta$. This is simply because 
$\text{Br}(\mu\to e\gamma)$ is proportional to $(\delta a_\mu)^2$. 
Thus, the branching ratio remains the same for fixed $\delta a_\mu$. 

From Fig.~\ref{fig:epsCP-Brmeg}, in order to obtain $\delta a_\mu > 
2.1 \times 10^{-9}$ (1$\sigma$), the leptogenesis scenarios with 
$\epsilon_1 > 10^{-5}$ provide the generic lower bound of 
$\text{Br}(\mu\to e\gamma) > 10^{-13}$ for the hierarchy $M_2/M_1 
> 10$ with $M_1 \gsim 1.5 \times 10^{11}$GeV. This is 
the sensitivity at which we expect to observe $\mu\to e\gamma$ 
in near future such as in the MEG experiment\cite{MEG}. For smaller 
$\epsilon_1 = 10^{-6}$ with $M_2/M_1 = 10$, there is the parameter 
region where we will not detect $\mu \to e \gamma$, while for larger 
$M_2/M_1$ such as $= 100$, the lower bound exceeds the 
experimental sensitivity with $\delta a_\mu > 2.1 \times 10^{-9}$. 
On the other hand, for a smaller hierarchy case, $M_2/M_1 = 3$, 
the lower bound of $\text{Br}(\mu\to e\gamma)$ reaches $O(10^{-13})$ 
when $\epsilon_1$ is larger than $10^{-4}$, which corresponds to 
$M_1 \gsim 10^{12}$GeV. As a result, since a lot of leptogenesis 
models practically require large CP asymmetry, we expect to observe 
the lepton-flavour violating muon decay in near future in the light 
of the muon $g-2$ anomaly when $M_2/M_1$ is hierarchical. 

\section{Conclusions and Discussion}

The reported discrepancy of the muon $g-2$ has impacts on 
low-energy phenomena. In this letter, we focused on the lepton 
flavour violation. We showed that in the light of this anomaly, 
successful leptogenesis scenarios prefere a sizable branching 
ratio of the lepton-flavour violating muon decay. Since 
the muon $g-2$ anomaly favors superparticles to stay in rather 
low-energy regime, e.g., $\lsim 1$TeV, the SUSY contributions 
to the muon $g-2$ can be checked directly by probing the those 
particles in the forthcoming Large Hadron Collider. When the 
SUSY contributions will be confirmed in the experiment, neglecting 
resonant leptogenesis and setting $\mu_X > \mu_R$, we found 
that leptogenesis scenarios generically predict $\text{Br}(\mu\to 
e\gamma)$ to become larger than $O(10^{-13})$ in wide 
parameter regions. Thus, we expect to detect $\mu \to e\gamma$ 
in the near future experiment. 

In the analysis, we have focused on the CP asymmetry parameter, 
$\epsilon_1$, to discuss leptogenesis. The lepton asymmetry depends 
on the thermal history of the universe as well, particularly on the 
production channels of the right-handed neutrinos. As a natural 
channel, they are produced in the thermal bath effectively when 
the reheating temperature exceeds the right-handed neutrino scale. 
In this thermal leptogenesis scenarios, the lepton asymmetry is 
roughly estimated as $Y_L \sim 10^{-2} \kappa \epsilon_1$, where 
$\kappa$ is the efficient factor which is determined by the initial 
abundance of the right-handed neutrino and the wash-out effects 
(see \cite{Buchmuller:2004nz}). In the absence of the initial 
right-handed neutrinos, $\kappa$ can be as large as $O(10^{-1})$, 
and thus $\epsilon_1 \gsim 10^{-6}$ is needed to explain the present 
measurements of BAU. This maximal $\kappa$ is achieved when 
a washout mass parameter, $\tilde{m}_1=(Y_N)_{1i}(M_i)^{-1}(Y_N^*)_{1i}
\langle H_u\rangle^2$, is tuned to be about 0.001eV. We checked 
that this condition can be realized on the lines of the lower bound 
of $\text{Br}(\mu\to e\gamma)$. On the other hand, 
since the efficiency factor decreases very quickly when $\tilde{m}_1$ 
differs from the maximal efficiency value, the efficiency factor easily 
takes smaller value. In practice, we need $\epsilon_1 \gsim 2 \times 
10^{-6}$\cite{Buchmuller:2004nz}, and then the lower bound of 
$\text{Br}(\mu\to e\gamma)$ is obtained to be larger than $O(10^{-13})$ 
for $M_2/M_1 > 30$ with $\delta a_\mu > 2.1 \times 10^{-9}$~\footnote{
  Since lager $\epsilon_1$ corresponds to heavier $M_1$, 
  a higher reheating temperature is required, and thus the 
  cosmological problem of the thermal gravitino production 
  tends to be severer. 
}. 
In this parameter region, we expect to measure the flavour-changing 
decay of the muon in near future. 
In another scenario, the right-handed neutrino may be produced 
non-thermally. In this case, the resultant lepton asymmetry strongly 
depends on physics in high-energy scale. Even in this case, the lepton 
asymmetry is proportional to the CP asymmetry parameter, $\epsilon_1$, 
and we obtain the lower bound of $\text{Br}(\mu\to e\gamma)$. 

If the lapton-flavour violations will not be observed in the future 
experiments, the absence of signals does not always exclude 
the leptogenesis scenarios as a source of BAU. In this letter, 
we imposed several conditions to obtain the lower bound. 
First of all, we assumed no accidental cancellations in the 
decay amplitude of the lepton-flavour violation. However, 
it is possible to suppress the process in some specific cases. 
Actually, $\text{Br}(\mu\to e\gamma)$ can be lower than the 
bounds in Fig.~\ref{fig:epsCP-Brmeg} when the right-handed 
neutrinos have a special flavour structure which suppresses 
$|(Y_N^{\dagger}Y_N)_{12}|$. Another cancellation may 
happen between the chargino and neutralino contributions, 
at some spots in the parameter space. 
Also, taking into account initial flavour-changing components 
of the soft SUSY breaking parameters at the mediation scale, 
cancellations may also happen because their contribution can 
destructively interfere with those from the right-handed neutrino 
Yukawa coupling. The second possibility is obtained by concerning 
the assumption of $\mu_X > \mu_R$. The charged-lepton sector 
receives the flavour-changing corrections through the renormalization 
evolutions only when the soft SUSY breaking effects are mediated 
before the right-handed neutrinos decouple. We have assumed 
that the mediation takes place at the GUT scale, while the 
gauge-mediated SUSY breaking scenarios generally have lower 
$\mu_X$. The scale dependence is just logarithmic and very weak 
when the messenger scale is far from the right-handed neutrino 
mass scale. However, if the messenger scale approaches 
very close to the right-handed neutrino scale, the lepton-flavour 
violating processes become suppressed\cite{Tobe:2003nx}. 
In those cases, the lepton-flavour violations will not be measured 
in near future even in leptogenesis scenarios with the muon $g-2$ 
anomaly. 

The lepton asymmetry may be enhanced compared with the evaluation 
in this letter. Actually, the right-handed neutrinos can have a (quite) 
degenerate mass spectrum. In the case of the hierarchical spectrum, 
the lepton asymmetry is obtained only from the decay of the lightest 
right-handed neutrino, while when the heavier right-handed neutrino 
mass becomes close to the lightest one, {\it i.e.}, $M_2\to M_1$, 
CP asymmetric decay of $N_2$ simultaneously contributes to the lepton 
asymmetry\cite{ResoEPS,ResoLepto}. Actually, Eq.~(\ref{eq:formula-epsCP}) 
diverges for $M_2\to M_1$, and taking into account $\epsilon_2$ in addition 
to $\epsilon_1$, the CP asymmetry is enhanced and becomes maximized 
when $|M_2-M_1|\sim \Gamma_{2,3}+\Gamma_1$\cite{ResoEPS}. 
As an another possibility, the flavour effect might affect the leptogenesis. 
This effect can modify the estimation of the CP asymmetry of the 
right-handed neutrino decay. For instance, even if the total CP 
asymmetry, $\epsilon_1$, is canceled, sufficient CP asymmetry 
is potentially produced by flavour effects\cite{EPS1ZERO}.
Anyway, all of these possibilities restrict the right-handed neutrino 
structure. Thus, we may observe distinct signatures, e.g., $\tau \to 
\mu\gamma$ and $\tau \to e \gamma$, in future experiments in the 
light of the muon $g-2$ anomaly. 
We will discuss these contents in future works.

\section*{Acknowledgment}

The authors thank to Wilfried Buchm\"uller and Alejandro Ibarra 
for useful discussions and comments.


\end{document}